\documentclass{llncs}

\usepackage{amsmath,amssymb,amsfonts}
\usepackage{cite}
\usepackage[english]{babel}
\usepackage[scaled]{beramono}
\usepackage{booktabs}
\usepackage[T1]{fontenc}
\usepackage{graphicx}
\usepackage[utf8]{inputenc}
\usepackage{listings}
\usepackage{microtype}
\usepackage{newtxtext,newtxmath}
\usepackage{nicematrix}
\usepackage[subrefformat=parens]{subcaption}
\usepackage[normalem]{ulem}
\usepackage{url}
\usepackage{xcolor}

\definecolor{color-bg}{HTML}{F6F8FA}
\definecolor{color-keyword}{HTML}{D73A49}
\definecolor{color-ident}{HTML}{005CC5}
\definecolor{color-string}{HTML}{032F62}
\definecolor{color-comment}{HTML}{6A737D}

\begin{document}

\title{Performance Evaluation of a Next-Generation\\SX-Aurora TSUBASA Vector Supercomputer}

\author{Keichi Takahashi\inst{1}\orcidID{0000-0002-1607-5694} \and
        Soya Fujimoto\inst{2} \and
        Satoru Nagase\inst{2} \and
        Yoko Isobe\inst{2} \and
        Yoichi Shimomura\inst{1} \and
        Ryusuke Egawa\inst{3}\orcidID{0000-0001-8966-867X} \and
        Hiroyuki Takizawa\inst{1}\orcidID{0000-0003-2858-3140}}
\institute{Tohoku University\\
           \email{\{keichi,shimomura32,takizawa\}@tohoku.ac.jp} \and
           NEC Corporation\\
           \email{\{s-fujimoto,s.nagase,y-isobe-pi\}@nec.com} \and
           Tokyo Denki University\\
           \email{egawa@mail.dendai.ac.jp}}

\maketitle

\begin{abstract}
    Data movement is a key bottleneck in terms of both performance and energy
    efficiency in modern HPC systems. The NEC SX-series supercomputers have a
    long history of accelerating memory-intensive HPC applications by
    providing sufficient memory bandwidth to applications. In this paper, we
    analyze the performance of a prototype SX-Aurora TSUBASA supercomputer
    equipped with the brand-new Vector Engine (VE30) processor. VE30 is the first
    major update to the Vector Engine processor series, and offers
    significantly improved memory access performance due to its renewed memory
    subsystem. Moreover, it introduces new instructions and incorporates
    architectural advancements tailored for accelerating memory-intensive
    applications. Using standard benchmarks, we demonstrate that VE30
    considerably outperforms other processors in both performance and
    efficiency of memory-intensive applications. We also evaluate VE30 using
    applications including SPEChpc, and show that VE30 can run real-world
    applications with high performance. Finally, we discuss performance tuning
    techniques to obtain maximum performance from VE30.
\end{abstract}


\section{Introduction}

The \textit{memory wall} is a longstanding challenge in HPC that refers to the
continuously widening gap between arithmetic computing performance and memory
performance in a computing system. Due to the memory wall problem,
memory-intensive applications are bottlenecked by data movement and unable to
fully utilize the arithmetic computing performance of a system. Not only does
this hurt the performance of applications, but it also degrades energy
efficiency. The HPC community has therefore been actively exploring novel
architectures to tackle the memory wall, such as adopting high-bandwidth
memory devices for off-chip memory~\cite{Sato2020,Choquette2021}, implementing
large amounts of on-chip memory~\cite{Rocki2020,Louw2021}, and reducing
memory accesses by directly exchanging data between processing
elements~\cite{Hsu2021,Jouppi2021}.

However, these exotic architectures completely differ from general-purpose
CPUs, and often require the programmer to become familiar with unconventional
programming models. For example, the device could require multiple magnitudes
larger degree of parallelism than a CPU, or data movement across the memory
subsystem might require explicit management by the programmer. As a
consequence, developing software for such emerging hardware is generally
time-consuming and expensive. If a large body of users exists, the cost for
developing optimized software for a specialized system could be amortized
(\textit{e.g.,} deep learning). However, it is often the case in HPC that
a scientifically important software package is maintained by a handful of
programmers and used by a small group of users. In such a case, the
development cost becomes prohibitive.

NEC's SX-Aurora TSUBASA (SX-AT) supercomputer aims to achieve both
world-class memory performance and high productivity by a unique combination
of latest memory technology with the vector architecture. The vector
architecture has a long history and recently
regained interests from the community. This trend can be seen in the ARM Scalable
Vector Extension~\cite{Stephens2017} and upcoming RISC-V Vector
Extension~\cite{RISC-VFoundation2021}, both of which are heavily inspired by
the vector architecture. Since most HPC applications exhibit high data-level
parallelism that can be automatically exploited by a vectorizing compiler,
conventional software targeted for general-purpose CPUs can run with minor
modifications. To keep feeding data to the high-performance vector cores,
High Bandwidth Memory (HBM) is tightly coupled with the processor. As a result,
SX-AT offers massive memory performance to applications while ensuring
programmer productivity.

NEC has recently been prototyping a brand-new vector processor named Vector
Engine 3.0 (VE30) for SX-AT\@. VE30 takes a big leap from the previous Vector
Engine series, and brings a number of architectural advancements beyond peak
compute and memory performance increase. Specifically, VE30 introduces
bypassable per-core private L3 caches as a new level in the memory hierarchy
to accelerate cache-intensive applications. In addition, a new instruction
that performs indirectly addressed vector accumulation within a
compute-capable LLC is added.

Since these combined improvements are expected to accelerate applications
beyond the improvement of peak performance, application performance cannot be
trivially estimated. We therefore carry out the first performance analysis of
a next-generation vector supercomputer based on the VE30 processor. The main
contributions of this paper are summarized as follows.

\begin{itemize}
    \item This is the first work to evaluate the performance of a
        next-generation vector supercomputer equipped with VE30 processors.
        Using industry-standard benchmarks and several applications, we assess
        the real-world performance as well as the basic performance of VE30.
    \item This paper analyzes the performance gain obtained by each
        architectural improvement newly introduced in VE30 using
        microbenchmarks.
    \item This paper discusses performance tuning techniques to take
        advantages of the new architectural capabilities of VE30 to accelerate
        application performance.
\end{itemize}

The rest of this paper is organized as follows. Section~\ref{sec:ii}
introduces the NEC SX-AT supercomputer and describes the basic
architecture of the VE30 processor. Section~\ref{sec:iii} extensively
evaluates the performance of VE30 using standard benchmarks, microbenchmarks
and real-world workloads. Section~\ref{sec:iv} discusses performance tuning
techniques to fully exploit the potential of the VE30 processor.
Section~\ref{sec:v} concludes this paper.

\section{Overview of SX-Aurora TSUBASA VE30}\label{sec:ii}

In this section, we first outline the architecture of the SX-AT supercomputer,
and introduce the newly developed VE30 processor. We then describe the
architectural enhancements of VE30 from its predecessor.

\subsection{The SX-Aurora TSUBASA product family}

The \textit{SX-Aurora TSUBASA (SX-AT)} is the latest product family in the NEC
SX vector supercomputers series.  While SX-AT inherits the well-established
and successful design philosophy of its predecessors, it also embraces the
current \textit{de facto} standard HPC software ecosystem. The
first-generation SX-AT based on the Vector Engine 1.0 (VE10) processor was
released in 2018~\cite{Komatsu2018}, which was followed by the
second-generation SX-AT based on the Vector Engine 2.0 (VE20) processor
released in 2020~\cite{Egawa2020}. The third-generation SX-AT based on the
Vector Engine 3.0 (VE30) processor, which is evaluated in this paper, is
currently under development and will be released in the near future.

SX-AT employs a heterogeneous architecture consisting of a
\textit{Vector Host~(VH)} and a \textit{Vector Engine~(VE)}. The VH is an
x86 server responsible for running the OS and performing tasks such
as process and memory management and I/O. The VE is a vector processor
implemented on a PCI Express (PCIe) card, and executes the application.
The VH communicates with the VE over the PCIe link and controls the VE\@.

Although on the surface a VE resembles an accelerator such as a GPU, its
execution model differs substantially from that of a conventional
accelerator. Applications are fully executed on the VE, and system calls are
forwarded to the VH and handled by proxy processes running on the VH\@. This
design eliminates kernel launch overhead and reduces data transfer found in
conventional accelerators. Furthermore, this design allows users to develop
their applications using standard MPI and OpenMP-based programming models, and
does not require any knowledge of a vendor-specific programming language or
framework.

\subsection{Basic Architecture of the VE30 Processor}\label{sec:ii-b}

\begin{figure}
\centering
\begin{minipage}{.5\textwidth}
\centering
\includegraphics[width=.97\textwidth]{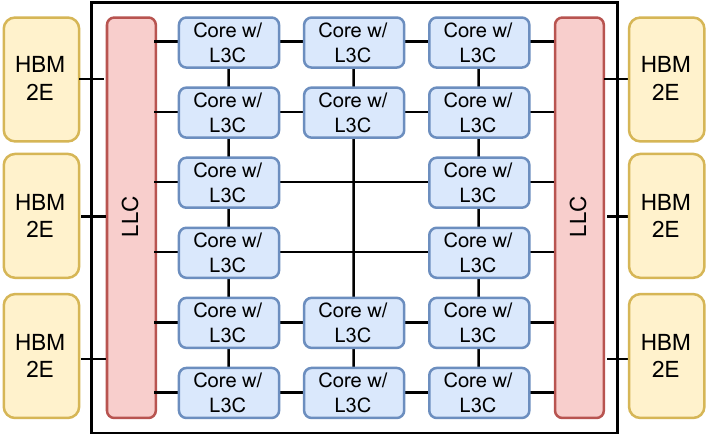}
\caption{Block diagram of the \\VE30 processor.}\label{fig:ve30-block}
\end{minipage}%
\begin{minipage}{.5\textwidth}
\centering
\includegraphics[width=.97\textwidth]{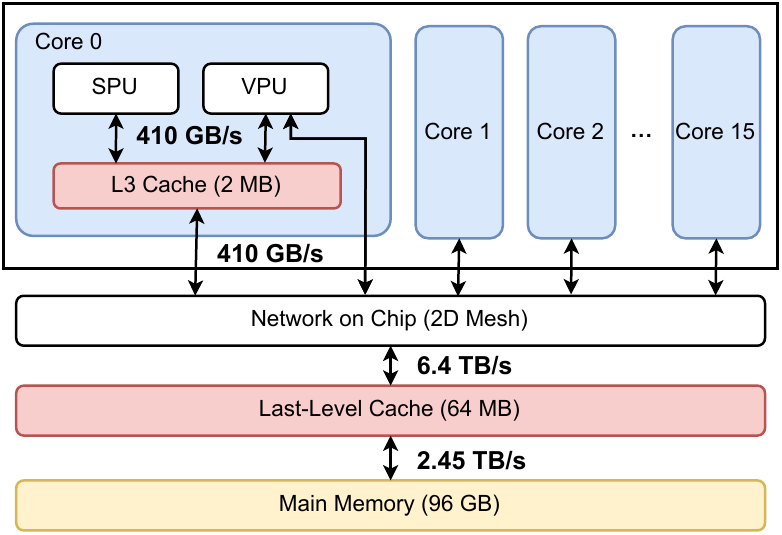}
\caption{Memory hierarchy of the \\VE30 processor.}\label{fig:ve30-memory}
\end{minipage}
\end{figure}

Figure~\ref{fig:ve30-block} illustrates an overview of a prototype VE30
processor, and Figure~\ref{fig:ve30-memory} depicts the memory hierarchy of a
VE30 processor. The VE30 processor integrates 16 vector cores, a shared LLC
and six HBM2E %
modules. Each vector core can perform up to
307.2\,GFLOP/s (DP) or 614.4\,GFLOP/s (SP), and thus a single socket performs
4.91\,TFLOP/s (DP) or 9.83\,TFLOP/s (SP) in total. The six HBM2E modules have
96\,GB of capacity and provide an aggregate memory bandwidth of 2.45\,TB/s to
the cores. The shared LLC is 64\,MB in size. The cores and LLC are
interconnected through a 2-dimensional Network on Chip (NoC).
VE30 provides a \textit{partitioning mode}, which splits
the cores, LLC and HBM in the processor into two NUMA nodes. This increases
the aggregate effective LLC bandwidth by alleviating congestion in the NoC,
and benefits LLC-intensive applications.

A vector core in VE30 comprises a Scalar Processing Unit (SPU) and a Vector
Processing Unit (VPU). The SPU fetches and decodes instructions, executes
scalar instructions and dispatches vector instructions to the VPU\@. An SPU
contains a 64\,KB L1 instruction cache, a 64\,KB L1 data cache, and a 512\,KB
unified L2 cache. A VPU contains 64 architectural vector registers that are
renamed to 188 physical vector registers. A single vector register holds up to
256 double-precision floating point elements (\textit{i.e.,} 2\,KB). A VPU
contains 32 vector pipelines, each of which has three Fused-Multiply Add (FMA)
execution units. Thus, in total, a vector core can perform 96 FMA operations
in a single cycle. The SPU and VPU share a 2\,MB unified L3 cache.

\subsection{Architectural Improvements from the VE20 Processor}

VE30 features a significantly advanced memory subsystem compared to its
predecessor.
First, the introduction of a new level in the memory hierarchy, per-core
private L3 caches, alleviates LLC contention and enables cache-intensive
applications to achieve higher performance. Second, the LLC capacity and
bandwidth are increased by 4$\times$ and $2.13\times$, respectively. Third,
both the capacity and bandwidth of the HBM are also improved. The peak HBM
bandwidth is increased by $1.60\times$ from 1.53\,TB/s to 2.45\,TB/s, and the
HBM capacity is doubled from 48\,GB to 96\,GB\@.
These drastic improvements to the memory subsystem combined are
expected to significantly accelerate both memory-intensive and cache-intensive
applications.

In addition to the enhancements made to the memory subsystem, the core count
is increased from 10 to 16 cores, which increases the peak single-socket
performance from 3.07\,TFLOP/s to 4.91\,TFLOP/s. It should be noted that,
despite the increase in the number of cores, the per-core cache and memory
performance is either increased or maintained.

Furthermore, a number of improvements are made to the core. First, VE30 relaxes
the alignment requirement for single-precision floating point vectors, and
improves the performance single-precision applications. Second, VE30
introduces a dedicated hardware mechanism for accelerating vector accumulation
with indirect addressing. These improvements do not directly contribute to the
peak FLOP/s rate, but are expected to benefit the performance of real-world
applications.

\footnotetext[1]{The peak performance is calculated based on the
AVX-512 Turbo Frequency when all cores are active.}
\footnotetext[2]{The LLC bandwidth (L2 bandwidth on IceLake-SP) is
measured using the Empirical Roofline Toolkit
(\url{https://bitbucket.org/berkeleylab/cs-roofline-toolkit}) since the peak
bandwidth is not disclosed by the manufacturers.}

\section{Performance Evaluation}\label{sec:iii}

In this section, we first reveal the basic performance of VE30 using
industry-standard benchmarks. We then use microbenchmarks to examine the
performance gains delivered by architectural improvements introduced
in VE30. Finally, we use workloads that represent practical applications to
assess the real-world performance of VE30. Note that the performance
measurements on VE30 are conducted using prototype software and hardware. Thus,
the results may be subject to change on the final product.

\subsection{Evaluation Environment}

\begin{table}
\centering
\caption{Specifications of the evaluated processors.}\label{tbl:processor-specs}
\scriptsize
\begin{NiceTabular}{l|l|l|l|l|l}
\toprule
                                                                           & VE Type 20B                                                 & VE Type 30A                                                 & A64FX                                                       & Xeon Platinum 8368                                                            & A100 80 GB PCIe                                                                                         \\ \midrule
Frequency [GHz]                                                            & 1.6                                                         & 1.6                                                         & 2.2                                                         & 2.4                                                                           & 1.412                                                                                                   \\
\begin{tabular}[c]{@{}l@{}}Performance\\ per Core [GFLOP/s]\end{tabular}   & \begin{tabular}[c]{@{}l@{}}307 (DP)\\ 614 (SP)\end{tabular} & \begin{tabular}[c]{@{}l@{}}307 (DP)\\ 614 (SP)\end{tabular} & \begin{tabular}[c]{@{}l@{}}70 (DP)\\ 140 (SP)\end{tabular}  & \begin{tabular}[c]{@{}l@{}}83.2 (DP)\footnotemark[1]\\ 166 (SP)\end{tabular} & \begin{tabular}[c]{@{}l@{}}181 (DP) w/ Tensor Core\\ 90 (DP) w/o Tensor Core\\ 181 (SP)\end{tabular}    \\
Number of Cores                                                            & 8                                                           & 16                                                          & 48                                                          & 38                                                                            & 108                                                                                                     \\
\begin{tabular}[c]{@{}l@{}}Performance\\ per Socket [TFLOP/s]\end{tabular} & \begin{tabular}[c]{@{}l@{}}2.4 (DP)\\ 4.9 (SP)\end{tabular} & \begin{tabular}[c]{@{}l@{}}4.9 (DP)\\ 9.8 (SP)\end{tabular} & \begin{tabular}[c]{@{}l@{}}3.3 (DP)\\ 6.7 (SP)\end{tabular} & \begin{tabular}[c]{@{}l@{}}3.1 (DP)\footnotemark[1]\\ 6.3 (SP)\end{tabular}  & \begin{tabular}[c]{@{}l@{}}19.5 (DP) w/ Tensor Core\\ 9.7 (DP) w/o Tensor Core\\ 19.5 (SP)\end{tabular} \\
LLC Bandwidth [TB/s]                                                       & 3.0                                                         & 6.4                                                         & 3.6                                                         & 3.2\footnotemark[2]                                                          & 4.9\footnotemark[2]                                                                                    \\
LLC Capacity [MB]                                                          & 16                                                          & 64                                                          & 32                                                          & 54                                                                            & 40                                                                                                      \\
Memory Bandwidth [TB/s]                                                    & 1.53                                                        & 2.45                                                        & 1.024                                                       & 0.204                                                                         & 1.935                                                                                                   \\
Memory Capacity [GB]                                                       & 48                                                          & 96                                                          & 32                                                          & 256                                                                           & 80                                                                                                      \\
Process Rule [nm]                                                          & 16                                                          & 7                                                           & 7                                                           & 10                                                                            & 7                                                                                                       \\ \bottomrule
\end{NiceTabular}
\end{table}

Table~\ref{tbl:processor-specs} summarizes the specifications of the
processors used in this evaluation. We compare VE30 to a variety of latest
processors used in HPC spanning from a vector processor, GPU, many-core
processor and general-purpose CPU\@: NEC Vector Engine Type 20B (an 8-core
SKU of VE20)~\cite{Egawa2020}, NVIDIA A100 40\,GB and 80\,GB PCIe
models~\cite{Choquette2021}, Fujitsu A64FX~\cite{Sato2020}, and Intel Xeon
Platinum 8368 (IceLake-SP)~\cite{Papazian2020}.
As shown in Table~\ref{tbl:processor-specs}, the peak performance of A100
doubles when the Tensor Cores are included. We use the peak performance
including the Tensor Cores when calculating the efficiency of HPL, and the
peak performance excluding the Tensor Cores for the other benchmarks.
This is because all benchmarks except HPL do not use the Tensor Cores.

Multi-node measurements for VE30 are carried out on a cluster composed of 16
VHs interconnected with a dual-rail InfiniBand HDR network. Each VH is
equipped with eight Vector Engine Type 30A cards, an AMD EPYC 7713P processor
and 512\, GB of DDR4-3200 SDRAM\@.

\subsection{Basic Benchmarks}

We use four widely recognized benchmarks in HPC to
evaluate the basic performance of VE30: the High Performance Linpack 
(HPL)~\cite{Dongarra2003} benchmark, STREAM benchmark, High Performance
Conjugate Gradients (HPCG)~\cite{Dongarra2016} and Himeno
benchmark~\cite{Himeno}.

HPL is a compute-intensive benchmark that solves a dense system of linear
equations using LU decomposition with partial pivoting. The STREAM benchmark
measures the effective memory bandwidth. HPCG is a memory-intensive benchmark
that solves a sparse linear system using the conjugate gradient method and a
geometric multigrid preconditioner. The Himeno benchmark is also
memory-intensive, and solves the Poisson equation using the Jacobi method.
Only the Himeno benchmark uses single-precision floating point numbers for
computation and the rest use double-precision floating point numbers. Since
HPL and HPCG executables optimized for the A64FX processor are unavailable to
us, the HPL and HPCG performance of A64FX is calculated based on the Top500
result of an A64FX-based system (\textit{Fugaku}~\cite{Sato2020}).

\begin{figure}
\centering
\begin{minipage}{.48\textwidth}
\centering
\includegraphics{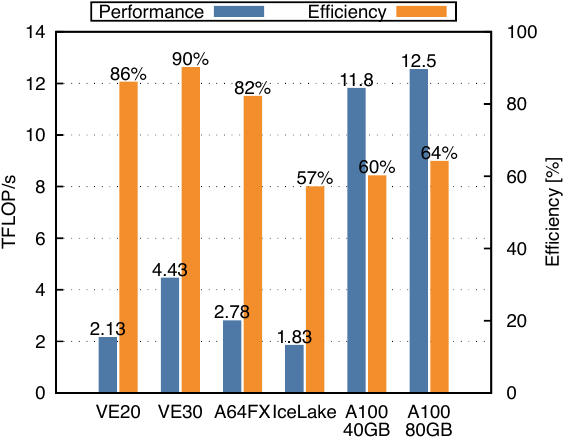}
\caption{HPL benchmark performance.}\label{fig:hpl}
\end{minipage}
\hspace{0.02\textwidth}
\begin{minipage}{.48\textwidth}
\centering
\includegraphics{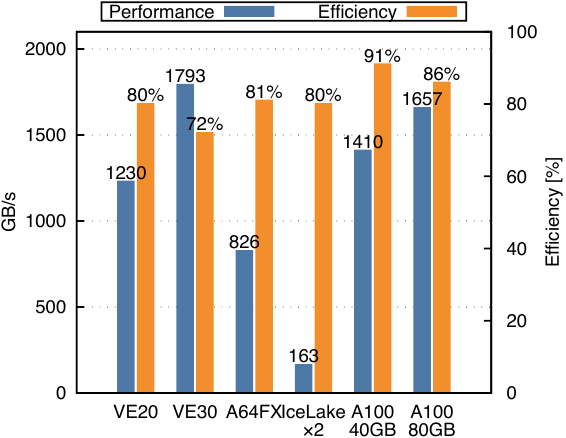}
\caption{Effective memory bandwidth.}\label{fig:stream}
\end{minipage}
\end{figure}

Figure~\ref{fig:hpl} compares the HPL performance of different processors.
The NVIDIA A100 clearly stands out from the other processors. The A100 40\,GB
model achieves over 11.8\,TFLOP/s in HPL performance, and the 80\,GB model
achieves a slightly higher performance of 12.5\,TFLOP/s due to the increased
problem size and higher TDP\@. The VE30 processor delivers 4.43\,TFLOP/s and
surpasses both A64FX and IceLake-SP\@. With respect to efficiency, VE30 is the
highest with an efficiency of 90\%, followed by A64FX and VE20. A100 shows
relatively low efficiency as it cannot maintain the GPU boost clock due to
power throttling.

Figure~\ref{fig:stream} compares the effective memory bandwidth of the
different processors measured using STREAM\@. The effective memory
bandwidth of VE30 exceeds 1.79\,TB/s and is clearly the highest among the
evaluated processors. Compared to its predecessor, VE20, the effective memory
bandwidth of VE30 is 1.45$\times$ higher.

\begin{figure}
\centering
\begin{minipage}{.48\textwidth}
\centering
\includegraphics{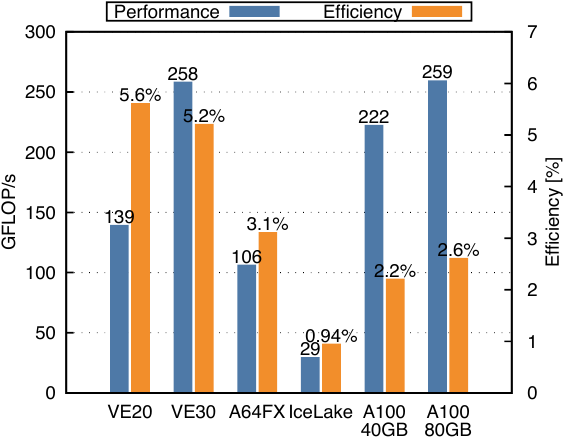}
\caption{HPCG benchmark performance.}\label{fig:hpcg}
\end{minipage}
\hspace{0.02\textwidth}
\begin{minipage}{.48\textwidth}
\centering
\includegraphics{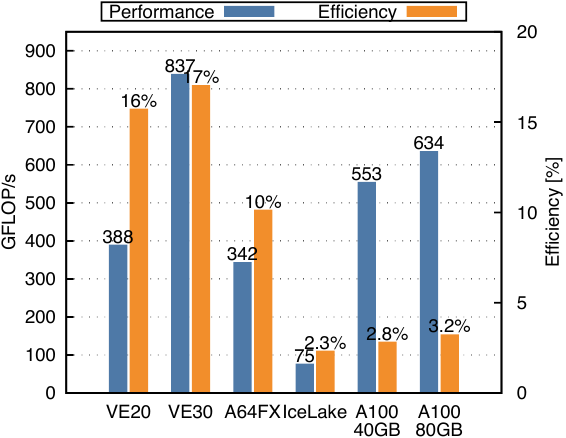}
\caption{Himeno benchmark performance (XL size).}\label{fig:himeno}
\end{minipage}
\end{figure}

Figure~\ref{fig:hpcg} shows the HPCG performance of the evaluated processors.
VE30 attains 258\,GFLOP/s and outperforms VE20, A64FX, IceLake-SP
and the A100 40\,GB model. It achieves almost identical performance as
the A100 80\,GB model. In terms of efficiency, VE30 achieves 5.2\% of the peak
performance, which is considerably higher than that of the other processors:
1.97$\times$ higher than the A100 80\,GB model and 5.72$\times$ higher than
IceLake-SP\@. Furthermore, VE30 achieves the highest energy efficiency among
all processors. The energy efficiency of VE30 when executing HPCG reaches
1.034\,GFLOP/s/W, while A100 40\,GB and 80\,GB models achieve 0.909\,GFLOP/s/W
and 0.999\,GFLOP/s/W, respectively. These results highlight that VE30
successfully strikes the balance between memory performance and floating-point
performance, whereas other processors heavily prioritize floating-point
performance over memory performance.

Figure~\ref{fig:himeno} shows the performance of the Himeno benchmark. VE30 is
the best-performing one among all processors. It marks 837\,GFLOP/s and
surpasses the A100 40\,GB and 80\,GB models by a factor of 1.51$\times$ and
1.32$\times$, respectively. Interestingly, the speedup exceeds the difference
in memory bandwidth. For example, VE30 has 1.27$\times$ higher memory
bandwidth than the A100 80\,GB model, but its performance is 1.32$\times$
higher. The speedup over VE20 that achieves 388\,GFLOP/s is 2.15$\times$, which
is again much larger than the 1.60$\times$ peak memory bandwidth improvement.
This is likely because the alignment restriction for single-precision vectors
is relaxed in VE30, and single-precision applications can be executed more
efficiently. 

\begin{figure}
\centering
\begin{minipage}{.57\textwidth}
\centering
\includegraphics{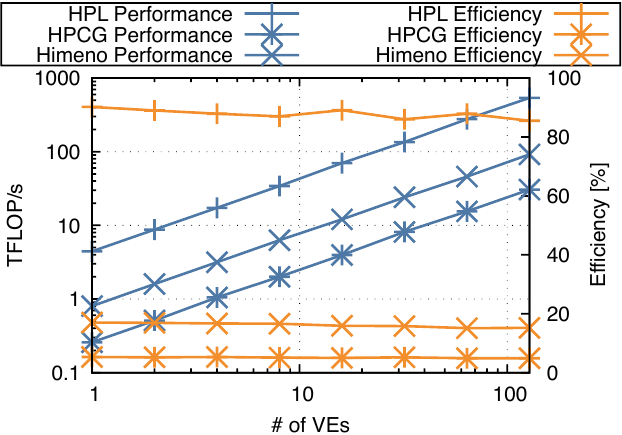}
\caption{Multi-node scaling performance of HPL, HPCG and Himeno benchmarks.}\label{fig:ve30-hpl-scalability}
\end{minipage}
\hspace{0.01\textwidth}
\begin{minipage}{.39\textwidth}
\centering
\centering
\captionsetup{type=table}
\caption{Summary of the Tohoku University kernel collection.}\label{tbl:isc-kernels}
\scriptsize
\begin{tabular}{@{}lll@{}}
\toprule
Kernel                              & Domain          & Bottleneck   \\ \midrule
Earthquake~\cite{Ariyoshi2007}      & Seismology      & Mem. B/W     \\
Turbulent Flow~\cite{Tsukahara2007} & Fluid dynamics  & LLC B/W      \\
Antenna~\cite{Sato2011}             & Electronics     & Mem. B/W     \\
Land Mine~\cite{Sato2003}           & Electronics     & Mem. B/W     \\
Turbine~\cite{Tsukahara2007}        & Fluid dynamics  & Mem. latency \\
Plasma~\cite{Katoh2005}             & Geophysics      & Mem. latency \\ \bottomrule
\end{tabular}
\end{minipage}
\end{figure}

Finally, we assess the multi-node scalability of HPL, HPCG and Himeno
benchmarks.
Figure~\ref{fig:ve30-hpl-scalability} shows the multi-node performance of the
two benchmarks as a function of the number of VEs. The results indicate that
all three benchmarks scale almost linearly from 1 VE to 128 VEs with minor
drop in efficiency. On 128 VEs, or 16 VHs, the HPL performance reaches 
537\,TFLOP/s with an efficiency of 85.5\%. The HPCG benchmark achieves 
30.6\,TFLOP/s on 128 VEs with an efficiency of 4.9\%. The Himeno benchmark
achieves 919\,TFLOP/s on 128 VEs with 15.2\% efficiency.

\subsection{Evaluation of Architectural Improvements}\label{sec:iii-c}

\subsubsection{Bypassable L3 Cache}\label{sec:iii-c-1}

VE30 incorporates per-core private L3 caches into the memory hierarchy. This
design choice was made based on the observation that cache-intensive
applications suffered from degraded LLC performance on previous generations of
the VE\@. This is largely due to the congestion in the NoC and cache
contention in the LLC\@. The introduction of private L3 caches is expected to
improve the effective cache bandwidth by alleviating NoC congestion and LLC
contention.

Furthermore, the L3 cache can be bypassed by software. Similar to non-temporal
loads and stores in CPUs and GPUs~\cite{Mittal2016}, each load or store
instruction can specify whether to bypass the L3 cache or not. Selectively
caching data that exhibit high temporal locality is expected to reduce cache
pollution and allow applications to efficiently utilize the limited cache
capacity. From the programmer's perspective, selective caching is
enabled by inserting a compiler directive \texttt{\#pragma \_NEC on\_adb(var)}
in the source code, where \texttt{var} indicates the array to be L3-cached.

Note that the L3 cache bypassing is different from the LLC retention
control~\cite{Onodera2021} that was available in the previous VE generations.
The LLC retention control allows applications to mark data as either temporal
or non-temporal when issuing loads and stores. The LLC then
prioritizes temporal data over non-temporal data when evicting cache lines.
However, even if an access is marked as non-temporal, it is still cached in
LLC\@. Thus, non-temporal data can still occupy a certain amount of the cache.
The L3 cache bypassing, on the other hand, completely bypasses the L3 cache.

To assess the contribution of the L3 cache to application performance, we
utilize the L3 cache bypassing feature and compare the performance of
applications with and without enabling the L3 cache. Here, we use the
\textit{Tohoku University kernel collection}~\cite{Soga2009,Komatsu2018}, a
set of computational kernels extracted from production applications developed
by the users of the Cyberscience Center, Tohoku University. As summarized in
Table~\ref{tbl:isc-kernels}, the kernel collection comprises six kernels
spanning a wide variety of scientific domains and performance characteristics.

\begin{figure}
\centering
\begin{minipage}{.62\textwidth}
\centering
\includegraphics{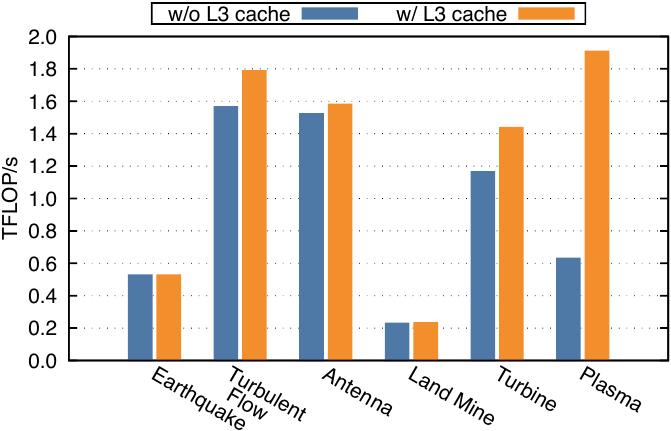}
\caption{Impact of L3 Cache on Tohoku University kernel collection performance.}\label{fig:isc-l3}
\end{minipage}
\hspace{0.02\textwidth}
\begin{minipage}{.34\textwidth}
\centering
\includegraphics{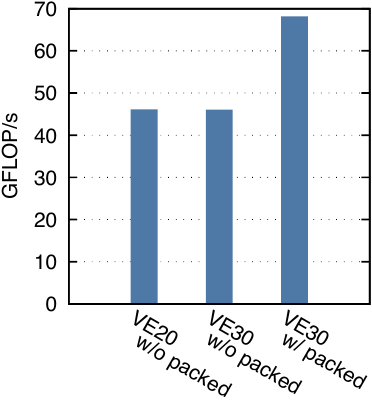}
\caption{Single-core performance of a single-precision 27-point stencil kernel.}\label{fig:packed}
\end{minipage}
\end{figure}

Figure~\ref{fig:isc-l3} presents the performance of each kernel with and without
enabling the L3 cache. The results reveal that Turbulent Flow, Antenna,
Turbine and Plasma clearly benefit from the L3 cache. Since the L3 cache
saves LLC and memory bandwidths by serving portion of the memory requests,
LLC-intensive and memory-intensive applications such as Turbulent Flow and Antenna
are accelerated. Contrastingly, Earthquake and Land Mine do not benefit
from the L3. These two kernels are memory-intensive and may either have poor
data locality or a large working set size that does not fit in the L3 cache.
The memory latency sensitive kernels, Turbine and Plasma, are also accelerated
as the L3 cache reduces memory latency. Accessing the LLC incurs higher
latency than the L3 cache since the LLC is physically farther away than the L3
cache, and requires communication over the potentially congested NoC.

\subsubsection{Relaxed Alignment Restriction for Packed Instructions}

The \textit{packed} instructions in SX-AT operate on vectors of
32-bit values, where each 64-bit element of a vector register holds a pair of
32-bit values. For example, a vector register can store 512 single-precision
floating point numbers using the packed format. However, previous generations
of the VE imposed an alignment restriction, requiring that the
starting address of a packed vector is 8-byte aligned. Otherwise, the packed
format cannot be used, and each element of a vector register holds only one
32-bit value instead of two. VE30 lifts this restriction and only requires
4-byte alignment for single-precision vectors.

To evaluate the speedup offered by the packed format, we measure the
performance of a single-precision 27-point stencil kernel.
Figure~\ref{fig:packed} presents the single-core
performance of the  27-point stencil kernel on VE20 and on VE30 with and
without using the packed format. Using the packed format on VE30 improves the
performance by 1.48$\times$ compared to VE20 and VE30 without using the packed
format.

\subsubsection{Hardware Support for Indexed Vector Accumulation}

The VE30 processor introduces hardware support for vector accumulation with
indirect addressing (\textit{i.e.,} \texttt{axpyi} in Sparse BLAS). Such
computation is fundamental in applications
including finite element and particle methods. An example of an indexed
vector accumulation is shown in Listing~\ref{lst:indirect}. In this example,
array \texttt{y} is indirectly accessed using array \texttt{l[i]} as
indices. This loop cannot be automatically vectorized by the compiler because
loop-carried dependencies exist if some of the indices in \texttt{l} overlap.

\begin{lstlisting}[caption={Indexed vector accumulation.},label={lst:indirect}]
for (int i = 0; i < n; i++)
    y[l[i]] = y[l[i]] + x[i];
\end{lstlisting}

Prior to VE30, programmers needed to manually examine whether \texttt{l[i]} may
overlap, and either insert the \texttt{ivdep} or \texttt{list\_vector}
compiler directive to the loop. The \texttt{ivdep} directive is specified when
there are no overlaps of indices, and simply vectorizes the loop. In the case
where the
indices do overlap, the \texttt{list\_vector} directive must be specified. The
\texttt{list\_vector} directive instructs the compiler to generate a code that
(1) computes the results using vector instructions ignoring loop-carried
dependencies, (2) checks the overlaps of indices, and (3) corrects the results
for overlapping indices using scalar instructions. However, the overhead
incurred by the corrections increases as the number of overlapping indices in
vector \texttt{l} increases.

VE30 adds specialized hardware for atomic accumulation in the LLC along with a
new instruction, \texttt{vlfa}. The \texttt{vlfa} instruction sends the vector
of indices (\texttt{l} in Listing~\ref{lst:indirect}) and the added vector
(\texttt{x}) to the LLC, and then performs the accumulation in the LLC\@. The
\texttt{vlfa} instruction should perform better than
\texttt{list\_vector} because scalar-based corrections are unneeded, and the
latency of vector gather is eliminated. Programmer productivity is also
improved since programmers no longer have to spend effort in identifying
whether the indices might overlap or not. Note, however, that \texttt{vlfa}
still slows down when the number of overlapping indices increases because
contention may occur in the LLC\@.

To investigate the performance of the \texttt{vlfa} instruction, we use the
indexed vector accumulation kernel shown in Listing~\ref{lst:indirect}. Here,
we compare the following five variants: scalar-only on VE20 and VE30,
\texttt{list\_vector} on VE20 and VE30, and \texttt{vlfa} on VE30. We vary the
number of overlapping indices to quantify the performance degradation caused
by the overlap of indices. This is achieved by initializing the index vector
\texttt{l} as the following:

\begin{equation}\label{eq:overlap}
\small
l[i] =
\begin{cases}
    0  & \text{if } i \mod 32 < k\\
    i  & \text{otherwise},
\end{cases}
\end{equation}
where $k$ is varied from 1 to 32.

\begin{figure}
\centering
\includegraphics{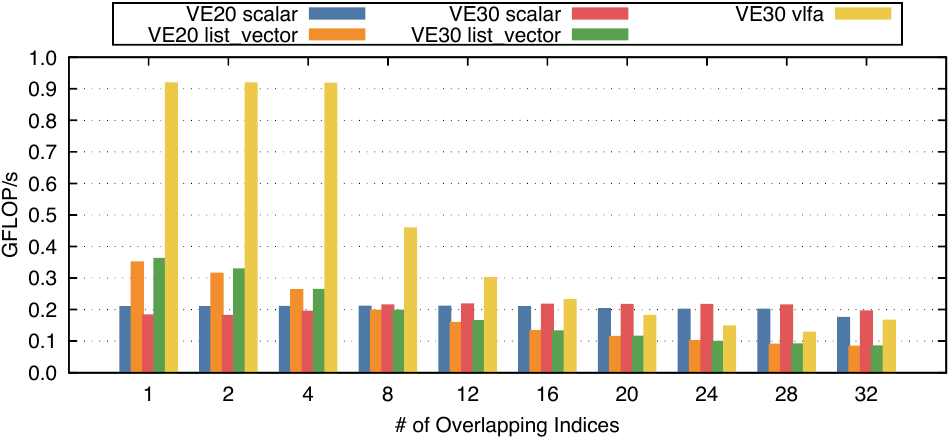}
\caption{Single-core performance of indexed vector accumulation.}\label{fig:vlfa}
\end{figure}

Figure~\ref{fig:vlfa} shows the performance of indexed vector accumulation for
each variant. Evidently, \texttt{vlfa} outperforms all other variants when the
number of overlapping indices is small; it is 3.48$\times$ faster than
\texttt{list\_vector} and 4.72$\times$ faster than scalar-only when 4 out of
32 indices overlap. The performance of \texttt{vlfa} starts to decline with
more than 8 overlapping indices, and falls below scalar-only with more than 20
identical indices. However, such large degree of address overlap is unlikely
in real-world applications. This indicates that programmers generally do not
need to specify the \texttt{ivdep} or \texttt{list\_vector} directives on VE30,
hence the productivity is improved.

\begin{lstlisting}[caption={A kernel loop involving indexed vector accumulation.},label={lst:cfd},float,language=fortran]
DO N = nstart,nend
  IF(flag3(N)==1) THEN
    COF(7,WI(N),WJ(N),WK(N))=COF(7,WI(N),WJ(N),WK(N))+W_TAUWC(N) * W_AREA_1(N)
    SOC(WI(N),WJ(N),WK(N))=SOC(WI(N),WJ(N),WK(N))+W_TAUWS(N) * W_AREA_1(N)
  ENDIF
ENDDO
\end{lstlisting}

Listing~\ref{lst:cfd} shows a kernel loop that appears in a real-world fluid
dynamics application. Here, two 3-dimensional arrays \texttt{COF} and
\texttt{SOC} are accumulated in a single loop. Prior to VE30, the VE compiler
was unable to vectorize this kind of code. On VE30, the compiler can vectorize
this code with the help of the \texttt{vlfa} instruction. As a result, this
kernel takes 175.6s to run on VE30 without the \texttt{vlfa} instruction, but
only takes 12.0s to run with the \texttt{vlfa} instruction, resulting in a
14.6$\times$ speedup.

\subsection{Real-world Workloads}

\subsubsection{SPEChpc 2021 Benchmark Suite}\label{sec:iii-d-1}

\textit{SPEChpc}~\cite{SPEChpc} is a benchmark suite developed by the Standard
Performance Evaluation Corporation (SPEC), and comprises a set of carefully
selected applications that represent a wide range of real-world HPC
applications. The latest version of the SPEChpc benchmark suite, SPEChpc 2021,
was released in October 2021. It supports multiple programming models and can
run on both CPUs and GPUs. In this evaluation, we use MPI+OpenMP on VE20, VE30,
A64FX and IceLake-SP, and MPI+OpenACC on A100.

We first use the \textit{tiny} workload from the SPEChpc 2021 benchmark suite
to compare the single-socket performance of the processors. 
The smallest tiny workload consists
of nine benchmarks and requires approximately 60\,GB of memory.
We plot the speedups to a reference system (a 2-socket
12-core Intel Haswell system) reported by the SPEChpc benchmark script for
each processor. If a processor needs more than one socket due to
the memory footprint requirement, the speedup is divided by the number of
sockets to make a fair comparison. Since the compilers for VE30 are still
under development as of writing this paper, we could not obtain the
performance results for SOMA and Minisweep on VE30.

\begin{figure}
\centering
\includegraphics{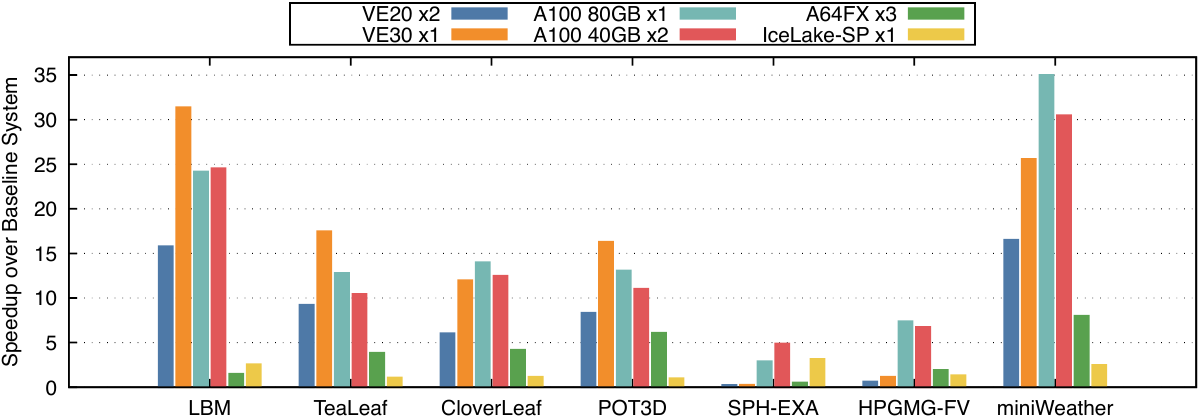}
\caption{SPEChpc 2021 tiny workload performance.}\label{fig:spechpc-tiny}
\end{figure}

Figure~\ref{fig:spechpc-tiny} summarizes the SPEChpc tiny workload performance
on the different processors. VE30 outperforms all other processors in LBM,
TeaLeaf and POT3D. The speedups of these three benchmarks over the A100 80\,GB
model are 1.29$\times$, 1.36$\times$ and 1.24$\times$,
respectively. The speedups of LBM and TeaLeaf exceed the difference in memory
bandwidth, suggesting that the architectural enhancements such as the newly
introduced L3 cache and increased LLC capacity and bandwidth, are contributing
to the performance gain.

VE30 also clearly outperforms A64FX and IceLake-SP in CloverLeaf and
miniWeather, but slightly underperforms the A100 40\,GB and 80\,GB models.
This is because the time-consuming kernels in CloverLeaf require a large
number of vector gather operations, and it appears that VE30 struggles at
hiding the latency of vector gather operations compared to A100. The
miniWeather benchmark contains a mix of memory-intensive and compute-intensive
kernels. Although memory-intensive kernels are faster on VE30 than on A100,
compute-intensive kernels are slower on VE30 and dominate the runtime.

SPH-EXA and HPGMG-FV perform poorly on VE30. SPH-EXA~\cite{Cavelan2020} is
mainly bottlenecked by the construction of an octree-based spatial index and
nearest neighbor queries over the index. Both of these functions inherently
require recursive function calls and cannot be vectorized. To achieve better
performance on vector processors, the nearest neighbor search needs to be
changed to a vector-friendly algorithm. 

HPGMG-FV suffers from short loop length. The HPGMG-FV tiny workload decomposes
a $512^3$ cubic domain into $32^3$ cubic boxes and distributes the boxes to
MPI ranks. The most time-consuming Gauss-Seidel Red-Black smoother kernel
sweeps over a box with a triple-nested loop each corresponding to a spatial
dimension. As a result, each loop runs for 32 times, but this is too short
compared to the vector length of a VE, which is 256 double-precision elements.
A potential optimization is collapse the nested loops and increase the loop
length. Another possible optimization is to offload the coarse grid levels to
the VH and process fine grid levels on the VE.

\begin{figure}
\centering
\includegraphics{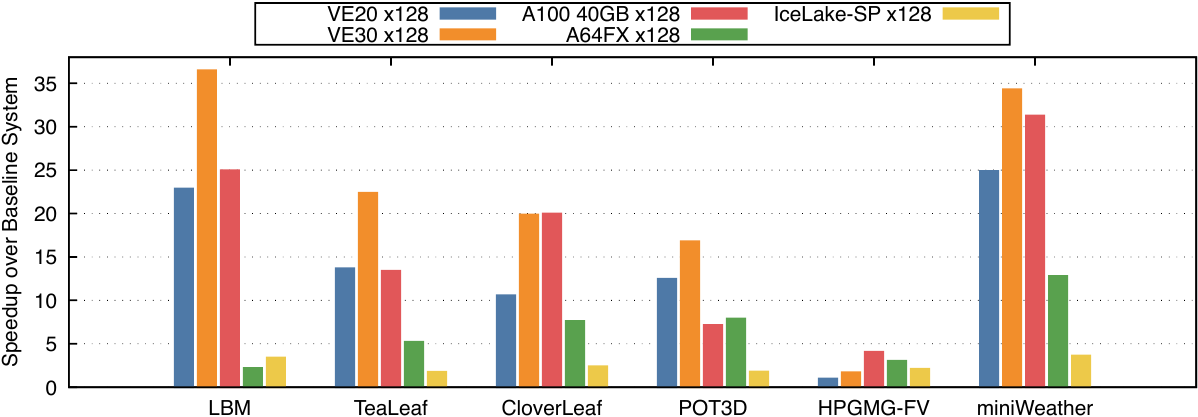}
\caption{SPEChpc 2021 medium workload performance.}\label{fig:spechpc-medium}
\end{figure}

To evaluate the multi-node scalability, we also compare the performance of the
SPEChpc \textit{medium} workload. The medium workload consists of six
benchmarks and requires approximately 4\,TB of memory. Here, we execute the
workload using 128 sockets on all processors. Results
for the A100 80\,GB model are unavailable since we do not have access to a
large-scale deployment of the A100 80\,GB model.
Figure~\ref{fig:spechpc-medium} summarizes the medium workload performance.
Here, VE30 is the fastest in four out of the six benchmarks, which are LBM,
TeaLeaf, POT3D and miniWeather. Compared to the tiny workload, the speedup of
VE30 over A100 is larger in the medium workload. For example, POT3D is
2.34$\times$ faster on VE30 compared to A100 in the medium workload, but it is
only 1.47$\times$ faster in the tiny workload. This suggests that VE30
provides better multi-node scalability than A100.

\subsubsection{Tohoku University Kernel Collection}

As described in Section~\ref{sec:iii-c-1}, the Tohoku University
kernel collection represents real-world applications developed by the users of
the Cyberscience Center at Tohoku University. Figure~\ref{fig:isc} shows the
performance of the Tohoku University kernels on VE20 and VE30.
Evidently, VE30 consistently outperforms VE20 with all kernels. The speedup is
especially significant for Turbulent Flow, Turbine and Plasma, all of which
perform more than 2.3$\times$ faster on VE30 than on VE20. Given that
Turbulent Flow is bound by LLC bandwidth on VE20, we believe the performance
gain is obtained from the 2.13$\times$ LLC bandwidth increase and the newly
added L3 cache. Turbine and Plasma benefit from the reduction in memory
latency thanks to the L3 cache as discussed in Section~\ref{sec:iii-c-1}.

\subsubsection{Rainfall-Runoff-Inundation Model}

The Rainfall-Runoff-Inundation (RRI) Model~\cite{RRI,Shimomura2022}
is a 2-dimensional numerical model that is  widely adopted in Japan to conduct
flood forecasts.
The governing equations are solved using the fifth-order Runge-Kutta method
with adaptive time step control. From the computational point of view, the
major kernels in the RRI model are memory-intensive, thereby suited for
execution on VEs.

\begin{figure}
\centering
\begin{minipage}{.62\textwidth}
\centering
\includegraphics{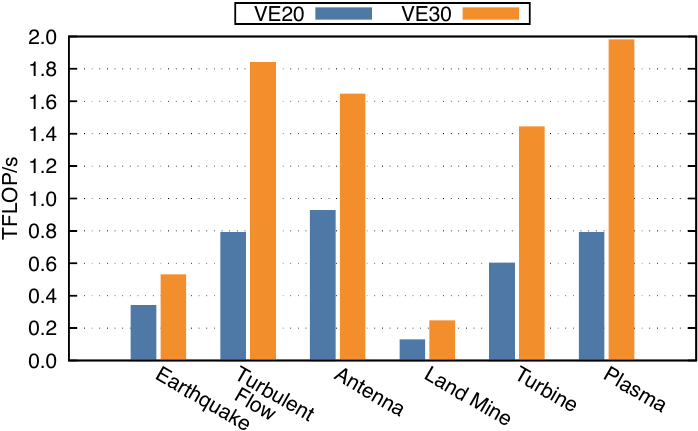}
\caption{Tohoku University kernel collection performance.}\label{fig:isc}
\end{minipage}
\hspace{0.02\textwidth}
\begin{minipage}{.32\textwidth}
\centering
\includegraphics{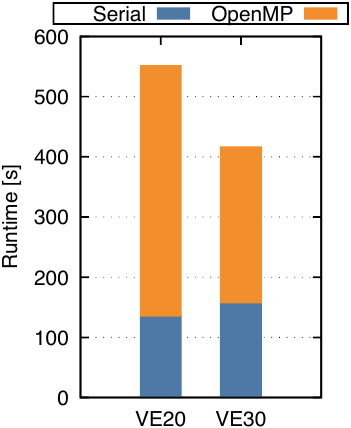}
\caption{Runtime of a 2-hour flood prediction using the RRI model.}\label{fig:rri_runtime}
\end{minipage}
\end{figure}

In this evaluation, we use an implementation of the RRI model optimized for
SX-AT, and measure the runtime required for conducting a 2-hour flood
prediction in the entire \textit{Tohoku} region of Japan.
Figure~\ref{fig:rri_runtime} shows the runtime of the flood prediction on VE20
and VE30. VE30 achieves 1.32$\times$ higher performance than VE20. The speedup
of the parallel regions is 1.60$\times$. Considering that the peak memory
bandwidth is increased by 1.60$\times$ from VE20 to VE30 and the RRI model is
memory-intensive, the observed speedup matches the expectation.

\section{Performance Tuning for VE30}\label{sec:iv}

Basic optimization techniques for VE include facilitating
vectorization by factoring out unvectorizable code from loops, increasing the
vector length using various loop transformations, and offloading
unvectorizable computation to the VH\@. In addition to these optimization
techniques, further performance can be exploited by utilizing the
architectural features introduced in VE30. In this section, we present such
tuning techniques and quantify their performance impact.

\subsection{Selective L3 Caching}

\begin{lstlisting}[caption={Jacobi method kernel in the Himeno benchmark.},label={lst:himeno},float]
for(i=1 ; i<imax-1 ; ++i)
  for(j=1 ; j<jmax-1 ; ++j)
    for(k=1 ; k<kmax-1 ; ++k){
      s0 = a[0][i][j][k] * p[i+1][j  ][k  ] + a[1][i][j][k] * p[i  ][j+1][k  ]
         + a[2][i][j][k] * p[i  ][j  ][k+1]
         + b[0][i][j][k] * ( p[i+1][j+1][k  ] - p[i+1][j-1][k  ]
                           - p[i-1][j+1][k  ] + p[i-1][j-1][k  ] )
         + b[1][i][j][k] * ( p[i  ][j+1][k+1] - p[i  ][j-1][k+1]
                           - p[i  ][j+1][k-1] + p[i  ][j-1][k-1] )
         + b[2][i][j][k] * ( p[i+1][j  ][k+1] - p[i-1][j  ][k+1]
                           - p[i+1][j  ][k-1] + p[i-1][j  ][k-1] )
         + c[0][i][j][k] * p[i-1][j  ][k  ] + c[1][i][j][k] * p[i  ][j-1][k  ]
         + c[2][i][j][k] * p[i  ][j  ][k-1] + wrk1[i][j][k];
      ss = ( s0 * a[3][i][j][k] - p[i][j][k] ) * bnd[i][j][k];
      wgosa += ss*ss;
      wrk2[i][j][k] = p[i][j][k] + omega * ss;
      // Copy wrk2 to wrk and sum wgosa across all ranks
    }
\end{lstlisting}

On VE30, programmers can take advantage of the bypassable L3 cache to
selectively cache frequently reused data. To demonstrate the effect of
selective L3 caching, we use the Himeno benchmark as an example.
Listing~\ref{lst:himeno} shows the time-consuming Jacobi kernel in the Himeno
benchmark. Arrays \texttt{a}, \texttt{b}, \texttt{c}, \texttt{wrk1} and
\texttt{bnd} are accessed in a consecutive manner and not reused. On the other
hand, array \texttt{p} is accessed in a stencil-like manner. Although
ideally 18 out of 19 accesses to \texttt{p} should hit in cache, the accesses
to the other arrays pollute the cache and degrade the cache hit ratio of
\texttt{p}. This cache pollution can be mitigated by caching \texttt{p} only
and bypassing the cache when accessing \texttt{a}, \texttt{b}, \texttt{c},
\texttt{wrk1} and \texttt{bnd}.

\begin{figure}
\centering
\begin{minipage}{.47\textwidth}
\centering
\includegraphics{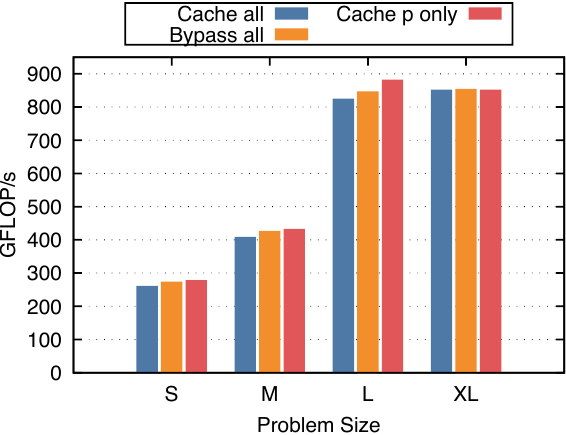}
\caption{Performance of Himeno benchmark with different problem sizes and L3 caching policies.}\label{fig:l3}
\end{minipage}
\hspace{0.01\textwidth}
\begin{minipage}{.47\textwidth}
\centering
\begin{subfigure}[t]{0.44\linewidth}
\centering
\includegraphics{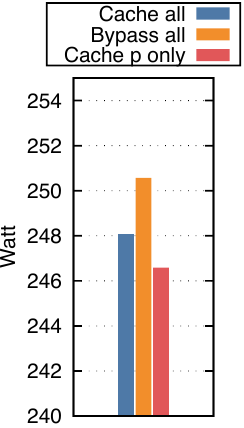}
\caption{Power Consumption}\label{fig:l3_power_consumption}
\end{subfigure}
\hspace{0.01\textwidth}
\begin{subfigure}[t]{0.44\linewidth}
\centering
\includegraphics{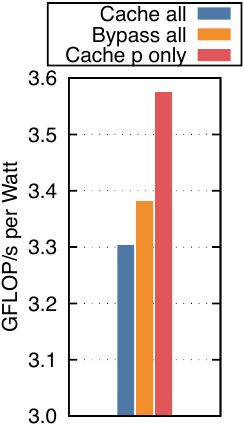}
\caption{Power Efficiency}\label{fig:l3_power_efficiency}
\end{subfigure}
\caption{Power efficiency of Himeno benchmark (L size).}\label{fig:l3_power}
\end{minipage}
\end{figure}

Figure~\ref{fig:l3} compares the performance of the Himeno benchmark under
three different caching policies: (1)~cache all arrays in the L3 cache,
(2)~always bypass the L3 cache, and (3)~only cache \texttt{p} in the L3 cache.
We also compare four different problem sizes: S ($2^6\times2^6\times2^7$), M
($2^7\times2^7\times2^8$), L ($2^8\times2^8\times2^9$), and XL
($2^9\times2^9\times2^{10}$). The results indicate that caching all arrays
does not show any notable improvement over bypassing all arrays. This suggests
that the L3 cache is polluted by non-temporal data and thus the cache hit
ratio of \texttt{p} is low. Selectively caching \texttt{p} improves
6.5\%, 5.7\% and 6.9\% over caching all arrays for problem sizes S, M and L,
respectively. This indicates that selective caching alleviates cache
pollution. Contrastingly, no performance improvement is observed for the XL
problem size. This is because \texttt{p} does not fit in the L3 cache in the
XL size.

To investigate if selective caching has an impact on power consumption and
power efficiency, we use the NEC Monitoring \& Maintenance Manager
(MMM)\footnote{\url{https://sxauroratsubasa.sakura.ne.jp/documents/guide/pdfs/InstallationGuide_E.pdf}} tool and measure the power consumption of the VE30 PCIe
card while running the Himeno benchmark.
Figure~\ref{fig:l3_power} compares the power consumption and efficiency
of the VE30 card under different caching policies. The
results indicate that selectively caching \texttt{p} reduces the power
consumption by 0.6\% compared to caching all arrays because the number of
memory
accesses is reduced. Combined with the performance improvement, selective
caching improves the power efficiency by 8.2\%, resulting in a power
efficiency of 3.57\,GFLOP/s/W. Compared to VE20 that achieves 2.21\,GFLOP/s/W
and the A100 40\,GB model that achieves 2.14\,GFLOP/s/W~\cite{Komatsu2021},
VE30 achieves 1.61$\times$ and 1.66$\times$ higher power efficiency,
respectively.

\begin{figure}
\centering
\begin{minipage}{.57\textwidth}
\centering
\begin{subfigure}[t]{0.31\linewidth}
\centering
\includegraphics{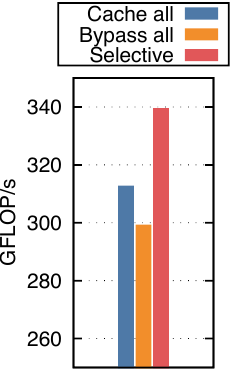}
\caption{Performance}\label{fig:lm_l3_perf}
\end{subfigure}
\hspace{0.01\textwidth}
\begin{subfigure}[t]{0.31\linewidth}
\centering
\includegraphics{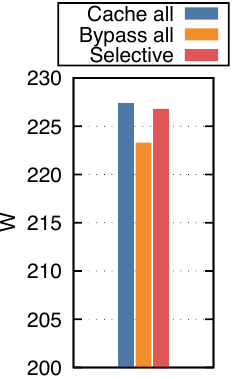}
\caption{Power Consumption}\label{fig:lm_l3_power}
\end{subfigure}
\hspace{0.01\textwidth}
\begin{subfigure}[t]{0.31\linewidth}
\centering
\includegraphics{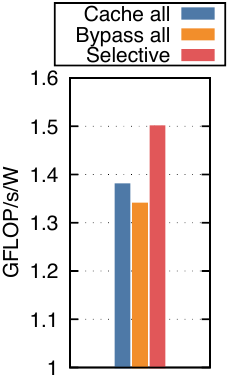}
\caption{Power Efficiency}\label{fig:lm_l3_power_eff}
\end{subfigure}
\caption{Performance and power efficiency of the Land Mine kernel under different L3 caching policies.}\label{fig:lm_pm}
\end{minipage}
\hspace{0.02\textwidth}
\begin{minipage}{.39\textwidth}
\centering
\includegraphics{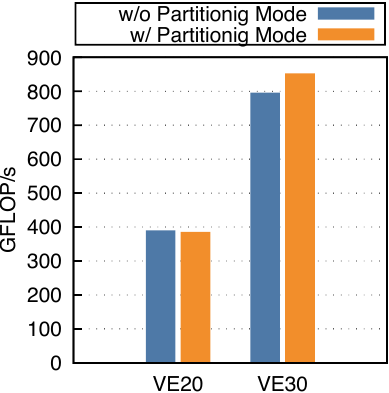}
\caption{Impact of the partitioning mode to Himeno benchmark performance (XL size).}%
\label{fig:himeno_pm}
\end{minipage}
\end{figure}

Furthermore, we apply selective L3 caching to the Land Mine kernel to study if
selective caching is beneficial for real-world applications. 
Figure~\ref{fig:lm_pm} shows the performance, power consumption and power
efficiency of the Land Mine kernel under different L3 caching policies. 
Bypassing the L3 cache yields the lowest performance of 299\,GFLOP/s. Enabling
the cache slightly improves the performance to 312\,GFLOP/s, and selective
caching further improves the performance to 339\,GFLOP/s. In terms of power
consumption, caching all arrays and selective caching both consume slightly
more power than bypassing the cache. This is because the increase in cache
power outweighs the reduction in memory power. However, the performance gain
of selective caching is large enough that its power efficiency is the highest.

\subsection{Partitioning Mode}

As mentioned in Section~\ref{sec:ii-b}, the partitioning mode
increases the effective LLC bandwidth by relieving the congestion in the NoC
that interconnects the cores and the LLC\@. Therefore, enabling the
partitioning mode may accelerate cache-intensive applications. Although the
partitioning mode has been available in the previous generations of VEs, its
benefits are expected to be larger on VE30 since NoC congestion becomes
heavier due to the increased number of cores.

To assess the effect of the partitioning mode on VE30, we measure the
performance of the Himeno benchmark with and without the partitioning mode.
The results are shown in Figure~\ref{fig:himeno_pm}. As expected, the
partitioning mode does not have a significant impact on VE20 since the NoC is
not congested. Contrastingly, the performance is increased by 7.1\% by
enabling the partitioning mode on VE30. This suggests that the NoC congestion
is alleviated by the partitioning mode. Thus, the use of partitioning mode
should be considered when running cache-intensive applications on VE30.

\section{Conclusions}\label{sec:v}

In this paper, we carried out an extensive performance evaluation of a
next-generation SX-AT supercomputer equipped with the brand-new VE30
processor. VE30 attains massive performance in memory-intensive standard
benchmarks such as the Himeno benchmark and outperforms other processors. The
speedup of VE30 over the other processors exceeds the difference in the peak
compute and memory performance, indicating the benefits of the novel
architectural enhancements introduced in VE30. VE30 also outperforms other
processors in many real-world applications such as SPEChpc. Finally, we
presented performance tuning techniques to fully exploit the potential of
VE30.

These evaluation results clearly demonstrate that VE30 can achieve high
sustained performance comparable to latest GPUs and CPUs, while allowing
programmers to use conventional programming models, \textit{i.e.}, MPI+OpenMP.
This proves the next-generation SX-AT to be an attractive choice for users
seeking real-world application performance.

\section*{Acknowledgments}

This work was partially supported by JSPS KAKENHI Grant Numbers JP20H00593,
JP20K19808, JP21H03449 and JP22K19764. Part of the experiments were carried
out using \emph{AOBA-A} and \emph{AOBA-C} at the Cyberscience Center, Tohoku
University, \emph{Flow} at the Information Technology Center, Nagoya
University, and \emph{SQUID} at the Cybermedia Center, Osaka University.

\bibliographystyle{splncs04}
\bibliography{references}

\end{document}